\documentclass[twocolumn,showpacs,preprintnumbers,amsmath,amssymb]{revtex4}

\usepackage{graphicx}
\usepackage{dcolumn}
\usepackage{bm}
\usepackage{multirow}

\begin{document}

\title{Observation of the Kohn anomaly near the K point of bilayer graphene}

\author{D. L. Mafra$^1$, L. M. Malard$^1$, S. K. Doorn$^2$, H. Htoon$^{2,3}$, J. Nilsson$^4$, A. H. Castro Neto$^5$, and M. A. Pimenta$^1$}
\address{$^1$Departamento de F\'{\i}sica, Universidade Federal de Minas
Gerais, 30123-970, Belo Horizonte, Brazil.\\
$^2$Chemistry Division, Los Alamos National Laboratory, Los Alamos,
New Mexico 87545, USA.\\
$^3$ Center for Integrated Nanotechnologies, Los Alamos National
Laboratory, Los Alamos, New Mexico 87545, USA.\\
$^4$Instituut-Lorentz, Universiteit Leiden, P.O. Box 9506, 2300 RA
Leiden, The Netherlands.\\
$^5$Department of Physics, Boston University, 590 Commonwealth
Avenue, Boston, Massachusetts 02215, USA.}

\date{\today}

\begin{abstract}

The dispersion of electrons and phonons near the K point of bilayer
graphene was investigated in a resonant Raman study using different
laser excitation energies in the near infrared and visible range.
The electronic structure was analyzed within the tight-binding
approximation, and the Slonczewski-Weiss-McClure (SWM) parameters
were obtained from the analysis of the dispersive behavior of the
Raman features. A softening of the phonon branches was observed near
the K point, and results evidence the Kohn anomaly and the
importance of considering electron-phonon and electron-electron
interactions to correctly describe the phonon dispersion in graphene
systems.

\end{abstract}

\maketitle

Graphene systems exhibit a strong electron-phonon coupling at
special points in the Brillouin zone, that softens the phonon energy
and gives rise to kinks in the phonon dispersion (infinities in
$\partial \omega/\partial \bold{q}$), which are called Kohn anomaly
\cite{piscanec04, Kohn}. This effect has been demonstrated
experimentally at the $\Gamma$ point of monolayer and bilayer
graphene using gated Raman scattering experiments
\cite{pisana07,yan07,das08,das09bi,yan08bi, malardprl08}. However,
the electron-phonon coupling is expected to be stronger at the K
point \cite{piscanec04}, but Raman experiments in graphene systems
performed with visible light cannot probe phonons near the K point
\cite{das08}. Moreover, in the case of AB-stacked bilayer graphene,
due to its special electronic and phonon structure
\cite{castronetoreview}, Raman experiments involve phonons closer to
the K point when compared to monolayer graphene
\cite{malardreports09}. This work presents a resonance Raman
investigation of AB-stacked bilayer graphene using many laser lines
in the near-infrared (near-IR) and visible range. The Kohn anomaly
for both symmetric (S) and anti-symmetric (AS) phonons was
evidenced, and results show the importance of considering higher
renormalization terms such as electron-electron interactions to
correctly describe the phonon dispersion near the K point
\cite{lazzeri08}. These effects are especially relevant for
understanding the dispersion of electrons and phonons
\cite{bostwick07,gruneisprl08,gruneis09} and transport properties in
this novel material \cite{castronetoreview}.

In a previous resonance Raman study of bilayer graphene performed in
the visible range \cite{lmalard07}, the electronic structure of
bilayer graphene was probed by analyzing the dispersion of
G$^{\prime}$ Raman band (also called 2D band) as a function of the
laser energy, and described within a tight-binding approximation
\cite{wallace47,McClure1957,SW1958} by determining the
nearest-neighbors hopping parameters $\gamma_0$, $\gamma_1$,
$\gamma_3$ and $\gamma_4$. It was shown in this work that a linear
iTO phonon dispersion provided a good fit of the experimental data
obtained with visible photons \cite{lmalard07}.

In the present work, we have extended the range of laser energies,
measuring the G$^{\prime}$ Raman band with many laser lines in the
range 1.33 to 2.81 eV. The measurements in the near-IR range (1.33
to 1.69 eV) are especially relevant since we can probe phonons that
are much closer to the K point. The analysis of the low energy data
allowed us to observe a non-linear softening of the phonon branch
near the K point, and the significant splitting of the symmetric (S)
and anti-symmetric (AS) phonon branches. In particular, we show that
the phonon softening is stronger for the S branch. Concerning the
electronic structure, we have also considered the in-plane
second-neighbor hopping parameter, which is expected to be of the
same order as the out-of-plane nearest-neighbor parameters, to
describe the G$^{\prime}$ Raman band dispersion.

The graphene samples used in this experiment were obtained by a
micro-mechanical exfoliation of graphite (Nacional de Grafite,
Brazil) on the surface of a Si sample with a 300 nm SiO$_2$
coverage. The laser power was kept at 1\,mW in order to avoid sample
heating. We used He-Cd, Ar-Kr and dye lasers for the laser lines in
the visible range (1.91--2.81 eV) and a Ti:Saphire laser for the
excitation in the near-IR range (1.33--1.69\,eV).

\begin{figure}
\includegraphics [scale=0.6]{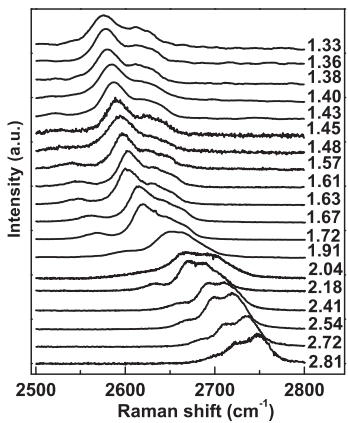}
\caption{\label{Fig1} The Raman G$^{\prime}$ band of bilayer
graphene recorded with 19 different laser lines between 1.33 and
2.81 eV (932 to 440 nm).} \label{fig1}
\end{figure}

Figure \ref{fig1} shows the G$^{\prime}$ band of bilayer graphene
recorded with 19 different laser lines between 1.33 and 2.81 eV (932
to 440 nm). We can see that both the position and the shape of the
G$^{\prime}$ band are strongly dependent on the energy of the
exciting laser. The G$^{\prime}$ band in graphene systems comes from
an intervalley double resonance (DR) Raman process
\cite{thomsen00,saito02} that involves one initial electronic state
with wavevector $k$ near the K point, one intermediate electronic
state with wavevector $k^{\prime}$ near the K$^{\prime}$ point, and
two in-plane transverse optical (iTO) phonons with wavevectors $q =
k + k^{\prime}$ \cite{malardreports09}. Since photons with different
energies excite electrons and phonons with different wavevectors $k$
and $q$, respectively, the dispersion of electrons and phonons near
the K point can be measured in when the energy of the incident
photons can be tuned \cite{malardreports09}.

In the case of AB-stacked bilayer graphene, the electronic branches
split into two valence bands ($\pi_1$ and $\pi_2$) and two
conduction bands ($\pi_1^*$ and $\pi_2^*$) \cite{castronetoreview}.
The iTO phonon branch also splits into two branches, related to the
symmetric (S) and anti-symmetric (AS) phonons. It has been shown
that the DR scattering process occurs preferentially in the high
symmetric $\Gamma$KM direction \cite{maultzsch04b,bob07}, involving
electrons in the K$\Gamma$ direction and phonons in the KM
direction.

Group theory analysis for bilayer graphene \cite{malardPRB09}
predicts four distinct DR processes (P$_{11}$, P$_{22}$, P$_{12}$
and P$_{21}$) along the $\Gamma$KM direction, which are illustrated
in Figs.\,\ref{fig2}(a) and (b). Since the lower and upper
conduction bands ($\pi_1^*$ and $\pi_2^*$) belong to different
irreducible representations, the S phonons (T$_{1}$ symmetry) are
associated with the P$_{11}$ and P$_{22}$ processes [see
Fig.\,\ref{fig2}(a)] involving electrons with same symmetry
($\pi_1^* \rightarrow \pi_1^*$ or $\pi_2^* \rightarrow \pi_2^*$)
whereas the AS phonons (T$_{2}$ symmetry) occur for processes
P$_{12}$ and P$_{21}$ [Fig.\,\ref{fig2}(b)] involving electrons with
different symmetries ($\pi_1^* \rightarrow \pi_2^*$ or $\pi_2^*
\rightarrow \pi_1^*$) \cite{lmalard07}.

\begin{figure}
\includegraphics [scale=0.42]{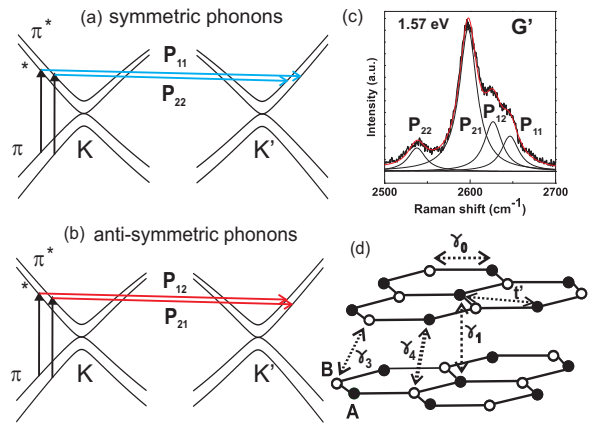}
\caption{\label{fig2} (Color online) (a) P$_{11}$ and P$_{22}$ DR
Raman processes involving the symmetric phonon, (b) P$_{12}$ and
P$_{21}$ DR Raman processes involving the anti-symmetric phonon (c)
G$^{\prime}$ Raman band of bilayer graphene measured with 1.57 eV
laser energy and fitted with four Lorentzian peaks with FWHM of 24
cm$^{-1}$. (d) Bilayer graphene atomic structure showing the hopping
processes associated with first-neighbor interactions ($\gamma_0$,
$\gamma_1$, $\gamma_3$ and $\gamma_4$), and second neighbor
interaction $t^{\prime}$.} \label{fig2}
\end{figure}

Each of these processes (P$_{11}$, P$_{22}$, P$_{12}$ and P$_{21}$)
is responsible for one peak in the G$^{\prime}$ band of  AB-stacked
bilayer graphene. Fig.\,\ref{fig2}(c) shows the Raman spectrum of
the G$^{\prime}$ band measured with 1.57\,eV laser energy and fitted
with four peaks. All have the same FWHM of 24\,cm$^{-1}$, which is
the FWHM of the single band in monolayer graphene
\cite{ferrari06,malardreports09}. We see in Fig.\ref{fig2}(a) that
the P$_{11}$ process involves the phonon with higher wavevector,
while the phonon with smaller wavevector gives rise to the P$_{22}$
process. Therefore, the P$_{22}$ process is especially relevant
since it involves phonons closer to the K point, even when compared
to the phonon in monolayer graphene probed by the same laser line.
Since the iTO phonon frequency increases with increasing $q$, we
conclude that the lowest and highest frequency peaks of the
G$^{\prime}$ band are associated with the P$_{11}$ and P$_{22}$
processes, respectively. The two intermediate Raman peaks are due to
the P$_{12}$ and P$_{21}$ processes, as shown in Fig.\ref{fig2}(c).

\begin{figure}
\includegraphics [scale=0.6]{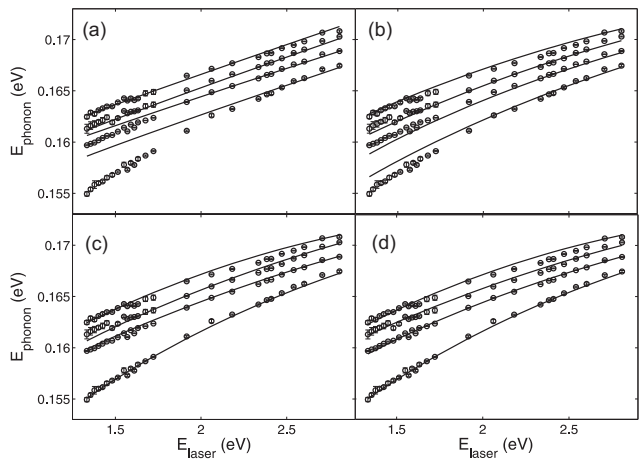}
\caption{\label{gamma4} Laser energy dependence of the peaks of the
G$^{\prime}$ bands, fitted considering four different
approximations. (a) Fitting considering linear phonon dispersion and
the SWM parameters $\gamma_0$, $\gamma_1$, $\gamma_3$, $\gamma_4$
(same approximation as in Ref. \cite{lmalard07}). (b) Fitting
considering the same non-linear phonon dispersion for the S and AS
branches, and the SWM parameters $\gamma_0$, $\gamma_1$, $\gamma_3$,
$\gamma_4$. (c) Fitting considering different non-linear phonon
dispersions for the S and AS phonon branches, and the same SWM as in
(b). (d) Fitting using te same phonon dispersions as in (c), the
first-neighbor SWM parameters $\gamma_0$, $\gamma_1$, $\gamma_3$,
$\gamma_4$, second-neighbor in-plane $t^{\prime}$ parameter, and the
$\Delta$ parameter, which represents the difference in energy of the
sublattices A and B. All fitting parameters used in approximations
(a), (b), (c) and (d) are shown in Table \,\ref{tab}.} \label{fig3}
\end{figure}

All G$^{\prime}$ bands shown in Fig.\ref{fig1} were fitted by four
Lorentzian curves, and laser energy dependence of the Lorentzian
peak positions are plotted in Fig.\ref{fig3}(a-d). In order to
analyze the experimental dispersion of the G$^{\prime}$ peaks shown
in Fig.\ref{fig3}(a-d), we need to consider the dispersion of both
electrons and phonons near the K point. The electronic dispersion
will be analyzed here using the tight-binding approximation, which
was first introduced by Wallace \cite{wallace47}, using the
Slonczewski-Weiss-McClure (SWM) \cite{McClure1957,SW1958} model for
graphite, and the phonon dispersion is obtained from the fit of the
experimental data.

In the previous resonance Raman study in bilayer graphene performed
in the visible range \cite{lmalard07}, a linear phonon dispersion
was considered to fit the G$^{\prime}$ peak positions versus
E$_{laser}$ data. Fig. \ref{fig3}(a) shows that the fitting of the
data of the present study considering the linear phonon dispersion
and the SWM parameters used in reference \cite{lmalard07} fails for
the experimental points in the near-IR region and, in particular,
for the data associated with the P$_{22}$ process involving phonons
closer to the K point.

In order to fit the low energy experimental data in Fig. \ref{fig3},
we have considered a non-linear relation for the iTO phonon
dispersion, given by a second-order polynomial $w(q)=A+Bq+Cq^2$.
Fig.\ref{fig3}(b) shows the fit considering the same non-linear
phonon dispersion for the S and AS phonon branches, and the
$\gamma_0$, $\gamma_1$, $\gamma_3$, $\gamma_4$ parameters.
Fig.\ref{fig3}(c) shows the fit using the same SWM parameters as in
Fig.\ref{fig3}(b) but considering two distinct non-linear phonon
dispersions for the S and AS branches. As we can see in
Fig.\ref{fig3}(b) and (c), the fitting in the low energy range is
improved considering the non-linear dispersion, but different
dispersions for the S and AS branches are needed to obtain a good
fit of the experimental data. All fitting parameters are shown in
Table \,\,\ref{tab}.

In the fittings shown in Figs. \ref{fig3}(a)-(c), we have considered
only the first-neighbor parameters $\gamma_0$, $\gamma_1$,
$\gamma_3$, $\gamma_4$. In principle, we could also introduce
higher-order terms and, in particular, the in-plane second-neighbor
parameter $t^{\prime}$, which is expected to be of the same order of
magnitude as the out-of-plane first-neighbor parameters. Figure
\ref{fig3}(d) shows the fit of the experimental data considering
also $t^{\prime}$ and $\Delta$, which represents the difference in
energy of the sublattices A and B. As we observe in Figure
\ref{fig3}(d), the resultant fit is slightly improved, mainly due to
the use of larger number of fitting parameters.

It is important to emphasize that slightly different values of the
$\gamma$ parameters are found when we include $t^{\prime}$ and
$\Delta$. A good fit can always be obtained for $\gamma_0$ values
ranging between 2.9 and 3.1 eV. Concerning the $\gamma_1$ parameter,
reasonable fits could only be obtained for $\gamma_1 <$ 0.35 eV, in
disagreement with values of $\gamma_1$ up to 0.4 eV proposed in the
literature \cite{IR1,IR2,IR3}. Considering the $\gamma_3$ parameter,
the best fit is obtained when $\gamma_3 \approx 0.1$ eV, and a
reasonable fit cannot be obtained for values of $\gamma_3>0.15$ eV.
Once again, this value is smaller than others found in the graphite
literature ($\gamma_3 \approx$ 0.30 eV
\cite{dresselhaus88,gruneis2008,schneider09}). Recent infrared
studies in exfoliated bilayer graphene consider $\gamma_3 \approx$
0.30 eV \cite{partoens06,min07}, but this value is not extracted
directly from the experiments \cite{IR1,IR3}. Notice that $\gamma_3$
is related to the trigonal warping effect (TWE) at very low energies
and gives rise to electron-hole pockets \cite{mccann2006a} in the
energy scale of $\sim$ 2 meV, which is not accessible in Raman
experiments. However as the energy increases, there is also a TWE
due to the symmetry of the lattice, which is probed by our
experiment. These two effects can explain the different $\gamma_3$
values found with distinct experimental techniques.

If we consider only nearest neighbor parameters to describe the
electronic structure, the best fit is obtained for $\gamma_4
\approx$ 0.15 eV, which is in close agreement with our previous
experiment \cite{lmalard07} and with Refs. \cite{IR1,IR2}. However,
smaller values of $\gamma_4$ provide a good fit when the
second-neighbor $t^{\prime}$ parameter is included. In fact, both
$\gamma_4$ and $t^{\prime}$ parameters are associated with the
asymmetry between electrons and holes in bilayer graphene. Finally,
reasonable fits can be obtained for different small positive and
negative values of $\Delta$ ($ | \Delta | < 0.01 $eV).

\begin{table*}
\centering \caption{Values of the SWM parameters (in units of eV)
and the iTO phonon dispersion parameters ($w(q)=A+Bq+Cq^2$) obtained
from the four different fits of the experimental data shown in
Fig.\ref{fig3} (a), (b), (c) and (d).}
 \label{tab}
  \centering
  \begin{tabular}{c|cccccccccccc}
    \hline
    {} & {} & {} & {} & {} & {} & {} & \multicolumn{3}{c}{symmetric} & \multicolumn{3}{c}{anti-symmetric} \\
    {} & $\gamma_0$ & $\gamma_1$ & $\gamma_3$ & $\gamma_4$ & $\Delta$ & $t^{\prime}$ & {A}(meV) & {B}(meV${\AA}$) & {C}(meV${\AA}^2$) & {A}(meV) & {B}(meV${\AA}$) & {C}(meV${\AA}^2$) \\
    \hline\hline\\
    (a) & \,\,\,\,2.9\,\,\, & \,\,\,\,0.3\,\,\, & \,\,\,\,0.1\,\,\, & \,\,\,\,0.12\,\,\, & - & - & \,\,\,\,153.7\,\,\, & \,\,\,\,38.5\,\,\, & \,\,\,\,-\,\,\, & \,\,\,\,154.0\,\,\, & \,\,\,\,38.8\,\,\, & \,\,\,\,-\,\,\, \\
    (b) & 3.0 & 0.35 & 0.1 & \,\,\,\,0.15\,\,\, & - & - & 149.3 & 69.5 & -46.6 & 149.3 & 69.5 & -46.6 \\
    (c) & 3.0 & 0.35 & 0.1 & 0.15 & - & - & 146.3 & 86.9 & -70.3 & 150.5 & 66.3 & -44.8 \\
    (d) & 3.0 & 0.35 & 0.1 & 0.10 & \,\,\,0.01\,\,\, & \,\,\,0.15\,\,\, & 146.3 & 86.9 & -70.3 & 150.5 & 66.3 & -44.8 \\
    \hline\hline
  \end{tabular}
\end{table*}

The analysis of the experimental data also gives us the phonon
dispersion for the symmetric (S) and anti-symmetric (AS) iTO phonon
branches near the K point. As shown in Figs.\,\ref{fig3}(c) and (d),
two distinct phonon branches, associated with symmetric and
anti-symmetric iTO phonons, are necessary to fit the experimental
data. The values of the polynomial parameters $A$, $B$ and $C$ for
the S and AS branches are shown in Table\,\ref{tab}. We observe that
the quadratic coefficients $C$ assume negative values, showing that
the slope of the phonon dispersion increases with decreasing $q$
values. This is direct evidence of the Kohn anomaly for the iTO
phonon branches, which are expected to exhibit a kink at the K
point.

\begin{figure}
\includegraphics [scale=0.4]{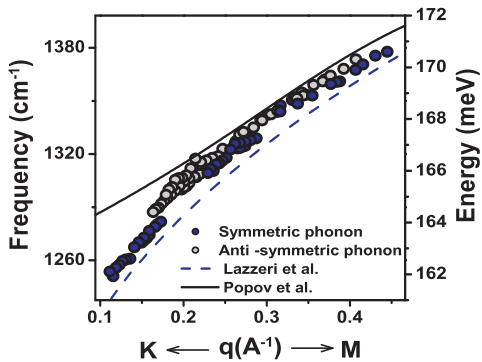}
\caption{(Color online) Experimental iTO phonon dispersion of
bilayer graphene near the K point of the S (full circles) and AS
(open circles) branches. The solid and dashed curves correspond,
respectively, to the theoretical iTO phonon dispersion of monolayer
graphene calculated using tight-binding by Popov \emph{et al.}
\cite{popov} and using DFT by Lazzeri \emph{et al.} \cite{lazzeri08}
within the $GW$ approximation.} \label{fig4}
\end{figure}

The experimental data shown in Fig. \ref{fig3} can be directly
plotted in a phonon energy dispersion relation ($\omega_{ph}$ versus
$q$) by eliminating the laser energy ($E_{laser}$) in the double
resonance conditions \cite{lmalard07}. Fig.\,\ref{fig4} shows the
phonon dispersion of the S (full circles) and AS (open circles)
branches obtained from the resonance Raman results, considering the
same TB parameters as in Fig.\ref{fig3}(c). Fig. \ref{fig4} also
shows the theoretical phonon dispersion near the Dirac point of the
iTO phonon branch of monolayer graphene calculated using the
tight-binding approximation by Popov \emph{et al.} \cite{popov}
(solid curve) and using DFT by Lazzeri \emph{et al.}
\cite{lazzeri08} (dashed curve) within the $GW$ approximation, where
electron-electron interaction is taken into account. Notice that the
phonon dispersion calculated by tight-binding fails to describe the
data for lower phonon energies, which are in good agreement with the
calculations within the $GW$ approximation. This result shows the
importance of considering electron-phonon and electron-electron
interactions in order to correctly describe the phonon dispersion
near the K point of graphene systems.

Another interesting observation is the fact that the phonon
softening is stronger for the S phonon branch as shown in
Fig.\,\ref{fig4}. This result is in agreement with the calculation
performed by Ando \emph{et al.} \cite{andoBi}, which predicts a
stronger phonon renormalization for the zone-center symmetric
phonon, due to distinct selection rules for interaction of S and AS
phonons with intra-valley and inter-valley electron-hole pairs. The
same type of selection rule is expected to occur for phonons near
the K point.

In summary, the dispersion of electrons and phonons of bilayer
graphene was investigated by performing a resonance Raman study of
the the G$^{\prime}$ Raman band, using laser energies from the
visible to the near-IR range. The measurements in the near-IR range
are especially relevant since we can probe phonons that are much
closer to the K point. The electronic structure was analyzed within
the tight-binding approximation, considering first and second
neighbor interactions. We have obtained accurate experimental data
for the phonon dispersion of the iTO branches near the K point, and
the phonon branch softening reveals the K point Kohn anomaly in
bilayer graphene. We have shown that the phonon renormalization is
stronger for the S phonon branch. Our results agree with the phonon
dispersion calculation which takes into account electron-electron
interaction in graphene systems, which plays an important role to
correctly describe the Kohn anomaly near the K point.

This work was supported by Rede Nacional de Pesquisa em Nanotubos de
Carbono - MCT, and the Brazilian Agencies CNPq and FAPEMIG.
Resonance Raman studies in the near infrared range were conducted at
the Center for Integrated Nanotechnologies, jointly operated by Los
Alamos and Sandia National Laboratories

\end{document}